%% file: main.tex
\documentclass[conference]{IEEEtran}

\usepackage[utf8]{inputenc}
\usepackage[T1]{fontenc}
\usepackage{graphicx}
\usepackage{todonotes}
\usepackage{listings}
\usepackage[2cell,matrix,all,rotate]{xy}
\usepackage{longtable}
\usepackage{tabularx}
\usepackage{xltabular}
\usepackage{booktabs}
\usepackage{soul}
\usepackage{censor}
\usepackage{cite}
\usepackage{amsmath,amssymb,amsfonts}
\usepackage{algorithmic}
\usepackage{graphicx}
\usepackage{textcomp}
\usepackage{xcolor}
\def\BibTeX{{\rm B\kern-.05em{\sc i\kern-.025em b}\kern-.08em
    T\kern-.1667em\lower.7ex\hbox{E}\kern-.125emX}}
\usepackage{caption}
\usepackage{paracol}

\makeatletter
\RequirePackage[bookmarks,unicode,colorlinks=true]{hyperref}%
   \def\@citecolor{blue}%
   \def\@urlcolor{blue}%
   \def\@linkcolor{blue}%

\def\orcidID#1{\smash{\href{http://orcid.org/#1}{\protect\raisebox{-1.25pt}{\protect\includegraphics{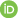}}}}}
\makeatother

 \lstdefinelanguage{Ecore}
 {
   morekeywords={
     package,
     class,
     property,
     ordered,
     composes,
     abstract,
     extends,
     attribute,
     operation,
     body,
     unique,
     and,
     or,
     forAll
   },
   sensitive=true, 
   morecomment=[l]{//}, 
   morecomment=[s][\bfseries\color{green!50!black}]{[}{]}, 
   morestring=[b]', 
   stringstyle=\color{blue}
 }
 
 \lstdefinelanguage{OclWithStrings}{
   language={Ocl},
   morekeywords={let,in,null, body,self},
   morestring=[b]', 
   stringstyle=\color{blue}
 }
 
 \lstdefinelanguage{Bytecode}{
   language={JVMIS},
   morekeywords={public,protected,private},
 }
 \newcommand{\java}[2][]{\lstinline[language=Java,postbreak={},basicstyle={\sffamily\small},#1]{#2}}
 
 \newcommand{\ecore}[1]{\lstinline[language=Ecore,postbreak={},basicstyle={\sffamily\small}]{#1}}

\begin{document}

	\title{{\huge MMT: Mutation Testing of Java Bytecode with Model Transformation -- An Illustrative Demonstration}}

\author{\IEEEauthorblockN{Christoph Bockisch \orcidID{0009-0006-9905-4798}}
\IEEEauthorblockA{
\textit{Philipps-Universität Marburg}\\
Marburg, Germany \\
}
\and
\IEEEauthorblockN{ Gabriele Taentzer \orcidID{0000-0002-3975-5238}}
\IEEEauthorblockA{\textit{Philipps-Universität Marburg}\\
Marburg, Germany  \\
}
\and
\IEEEauthorblockN{ Daniel Neufeld \orcidID{0009-0004-6697-5481}}
\IEEEauthorblockA{\textit{Philipps-Universität Marburg}\\
Marburg, Germany  \\
}
}

	\maketitle
	
	\input{0_abstract}

	\input{1_introduction}

	\input{3_approach}

	\input{4_tool_support}

	\input{6_related_work}

	\input{7_conclusion}

	\bibliographystyle{IEEEtran}
	\bibliography{bibliography}
	
	\pagebreak
	
	\input{appendix}

\end{document}

%% file: 0_abstract.tex
\begin{abstract}
Mutation testing is an approach to check the robustness of test suites.  
The program code is slightly changed by mutations to inject errors. 
A test suite is robust enough if it finds such errors.
Tools for mutation testing usually integrate sets of mutation operators such as, for example,  swapping arithmetic operators; modern tools typically work with compiled code such as Java bytecode.
In this case, the mutations must be defined in such a way that the mutated program still can be loaded and executed.
The results of mutation tests depend directly on the possible mutations. 
More advanced mutations and even domain-specific mutations can pose another challenge to the test suite.
Since extending the classical approaches to more complex mutations is not well supported and is difficult, we propose a model-driven approach where mutations of Java bytecode can be flexibly defined by model transformation.
The corresponding tool called MMT has been extended with advanced mutation operators for modifying object-oriented structures, Java-specific properties and method calls of APIs, making it the only mutation testing tool for Java bytecode that supports such mutations.
\end{abstract}


%% file: 1_introduction.tex
\section{Introduction}
\label{sec:introduction}

Mutation testing is a technique for testing the quality of a test suite. 
The assumption is that a program is well-tested if its tests can eliminate most of the errors injected into the program by mutations~\cite{offutt2001mutation}. 
The mutations to apply to the program are defined by mutation operators that describe syntactic changes to the program code.
Since the effectiveness of a mutation testing approach directly depends on its mutation operators' potential to find missing tests in the test suite, their variance and strength play a central role.

Mutations for Java programs can either be performed on the source code or the bytecode.
For both cases tools exist to support writing code transformations, for example, Polyglot~\cite{Nystrom2003} for source code, and ASM~\cite{asm} or BCEL~\cite{bcel}  for bytecode.
While mutating source code is often straightforward, this approach also has drawbacks:
Each mutated program must be recompiled, which can lead to significant compilation overhead~\cite{Jus14}. 
In addition, relevant information is often implicit in the source code; the absence of such information during mutation can easily lead to illegal code, and resolving this problem may require copying the functionality of a compiler.
An example is the invocation of overloaded methods, where the compiler performs a type analysis to select the proper implementation; in bytecode the method reference is already unique.
Therefore, newer mutation testing tools predominantly use bytecode mutation.

Looking at Java bytecode mutation tools like Major~\cite{Jus14}, Javalanche~\cite{schuler2009javalanche}, PITest~\cite{CLH+16},  they all provide mutation operators for manipulating arithmetic operators like replacing `+' with `-', relational operators like replacing `==' with `!=',  operators that replace return values, and operators that change the values of variables and constants. 
However, they do not provide mutation operators for manipulating the object-oriented structure, Java-specific properties of a program and library method calls, and therefore cannot assess the ability of a test suite to detect errors in the use of such concepts.

\begin{figure}[h]
  \begin{minipage}[t]{\columnwidth}
  \centering
    \includegraphics[width=\textwidth]{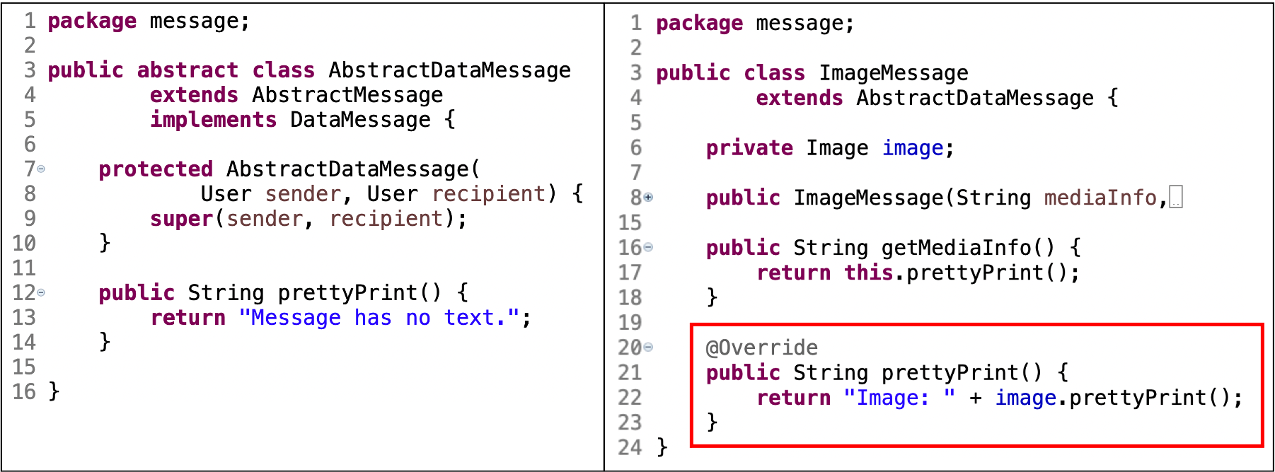}
   {\centering a) Developer-written application code.}
  \end{minipage}

  \begin{minipage}[t]{\columnwidth}
  \centering
    \includegraphics[width=\textwidth]{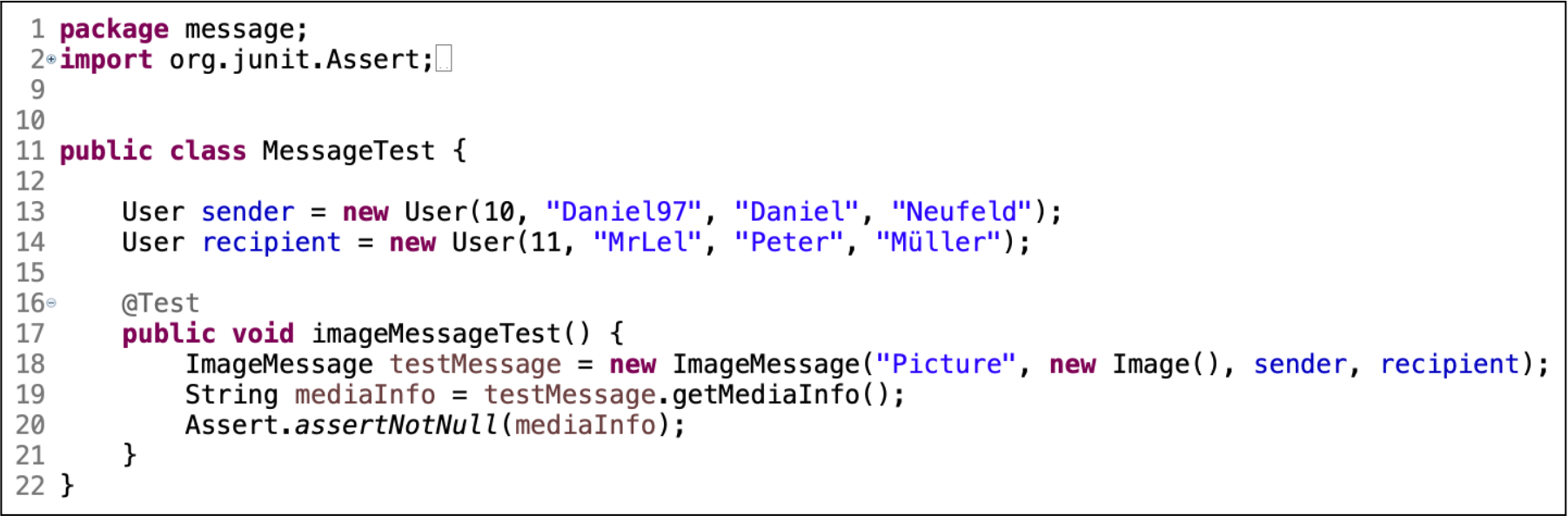}
    {\centering b) Developer-written test case code.}
  \end{minipage}
  \caption{Example code to which mutation testing can be applied. When the  code in the red box is removed, the test still succeeds.}
  \label{fig:init-example}
  \vspace{-.5cm}
\end{figure}

\begin{figure}
\centering
  \includegraphics[width=0.9\columnwidth]{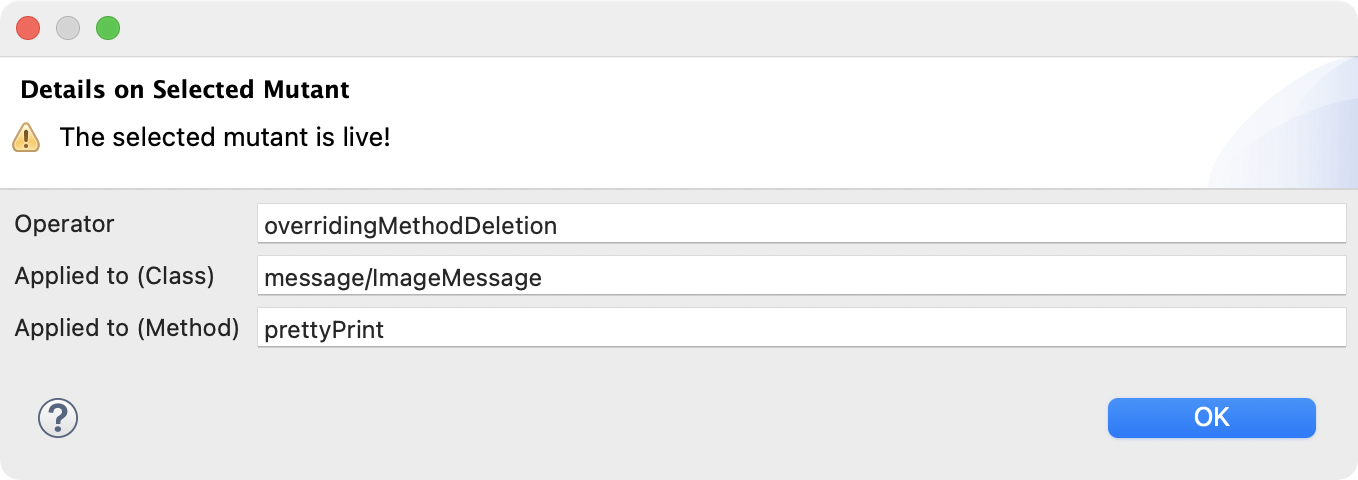}
  \caption{Details reported on a specific mutant, indicating a live mutant.}
  \label{fig:init-example-detail}
  \vspace{-0.5cm}
\end{figure}

\paragraph*{Example}
Fig.~\ref{fig:init-example} illustrates this technique with an example that is currently only supported by our tool among modern mutation testing tools for Java bytecode.
The figure shows a code snippet from a Java project (top) and a corresponding test case (bottom).
The test invokes the method \java{getMediaInfo} of \java{ImageMessage} which in turn returns the result of \java{prettyPrint}.
It is asserted that the method returns an actual String and not \java{null}.
One possible mutation is to remove the method \java{prettyPrint} from the class \java{ImageMessage} (marked by the red box).
The resulting code is still executable because a concrete implementation of the method \java{prettyPrint} is inherited from the superclass \java{AbstractDataMessage}.
While the result \java{"Message has no text."} is obviously undesired in the case of the mutant, the test is successful when run with both, the original version and the mutant.
Fig.~\ref{fig:init-example-detail} shows the details of the mutation that are reported by our mutation testing tool MMT for the described mutant.
It shows the kind of mutation, where it was applied and that it has not been detected by any test case, indicating that the test suite is too weak.
This conclusion is obviously correct, since the test does not actually inspect the result value.
Even, when trying to write a better assertion we discover that the method \java{prettyPrint} in \java{ImageMessage} is probably wrong as it includes the inappropriate dummy message returned by the super implementation.

One reason we believe that bytecode mutation testing tools do not provide advanced mutations is the complexity of bytecode consistency rules and the fact that such advanced program constructs may be spread across multiple locations in the bytecode; whether a method is private or not, for example,  affects all its call sites.
The bytecode manipulation tools mentioned above abstract from only some of these complexities. 
Consequently,  developers of mutation operators have to cope with a fixed, relatively low level of abstraction. 
If an easy-to-use approach were available for specifying mutation operators at different abstraction levels of the bytecode, the advantages of bytecode-level mutation would outweigh the complexity, and extending bytecode mutation operators to advanced program constructs would be feasible.
We believe that a model-driven approach can help here, and therefore propose our approach and tool called \emph{Model Mutation Testing} (MMT).

To allow the developer to choose the right level of abstraction, a model-driven approach for transforming Java bytecode was developed in \cite{BTNW20}. 
It builds on the Mod-BEAM approach~\cite{YBRA17}, where Java bytecode is represented as a model, along with a parser that reads in bytecode and creates a model, and a printer that creates bytecode from a model. 
In \cite{BTNW20}, the rules of the bytecode verifier (which is used by the Java Virtual Machine to identify inconsistencies comparable to compiler error) are specified as OCL constraints to provide a basis for automatically checking the syntactic correctness of program transformations.  
This will help us to specify mutation operators that generate only syntactically correct programs. 
The contributions of this paper are as follows:

\begin{itemize}
\item Our mutation testing tool MMT uses a model-driven approach to implement mutation operators, automatically applies such mutation operators, runs the test suite on the mutants, and summarizes the errors caused by the mutation that were missed by the test suite.

\item For an easy use of MMT, we provide it as an Eclipse plug-in that integrates with the graphical user interface of the Java Development Tools and provides full control over the process of mutation testing, allowing end users to control the set of mutation operators to be applied and the scope in which mutations should be applied.

\item MMT includes the standard set of mutation operators as well as a set of advanced mutation operators that modify the object-oriented program structure and change the use of the Java Collection Framework and other libraries; altogether 68 mutation operators.
Custom mutations can be implemented as model transformation rules or units in Henshin~\cite{ABJKT10} and freely added to MMT.
Since the Mod-BEAM meta-model abstracts some technical details from the bytecode, mutation operators can be implemented with minimal knowledge of Java bytecode.

\item A comparison of MMT with other mutation testing tools shows that it is the only tool for Java bytecode that provides advanced mutation operators. 
This wider range of mutation operators can be considered as a necessary prerequisite for generating diverse mutants. 

\end{itemize}

The paper is organized as follows: 
In Section~\ref{sec:approach}, we present our approach to mutation testing of Java bytecode, and 
 we give an overview of the tool support in MMT in Section~\ref{sec:toolsupport}. 
Section~\ref{sec:relatedwork} discusses the related work and Section~\ref{sec:conclusion} draws a conclusion.
The appendix contains a tool demo.

%% file: 3_approach.tex
\section{Approach}
\label{sec:approach}
In this section, we present our model-driven approach to mutation testing of Java bytecode.  
First, we give an overview of the mutation testing workflow  in general and using a model-driven approach. 
Then, we briefly recall the Mod-BEAM approach, which provides a model for Java bytecode.
Program mutation can thus be performed as model transformation.
Mutation operators should be developed so that they always provide syntactically correct and executable bytecode. 

\subsection{Workflow}
The concept of mutation testing can be used to measure the quality of a test suite for a given program.
Thereby, all tests must initially be successful, which is no limitation to generality as failing tests can be disabled.
A mutation testing tool will repeatedly mutate the program by injecting an error into the code.
Subsequently, the test suite is run on each of these mutated programs.
If there is a test that fails on the mutated program, that mutant is {\em killed}. 
Mutations that have not been detected are called {\em living}.

Next, the \emph{mutation score} is calculated by dividing the number of killed mutants by the number of generated mutants.
It serves as a measure of the quality of a test suite and is designed to help developers write more efficient tests, either by adding new test cases or by strengthening the assertions of existing tests to detect the living mutants.
A mutation score of 1.0 states that all mutants generated can be killed by the test suite. 

To mutate Java bytecode in our model-driven approach, it is first parsed and translated into a bytecode model.
Then, this model is transformed by applying a model transformation rule or, if the mutation is more complex, applying an entire transformation unit. 
The resulting model is translated back to bytecode.
We use the Mod-BEAM approach~\cite{BTNW20} based on the Eclipse Modeling Framework (EMF)~\cite{emfWebsite} to translate between bytecode and model, which is presented below. 
To specify and perform model transformations, any model transformation approach working on EMF models can be used in principle. 
We use Henshin~\cite{ABJKT10}, a model transformation language and environment for EMF, that is based on graph transformation concepts and supports formal validation. 

Defining mutations as model transformations in conjunction with the characteristics of Mod-BEAM leverages the extensibility of our approach with new mutation operators for several reasons.
Mutation operators can be written in a declarative manner, which makes them very readable.
Furthermore, all model elements can be accessed uniformly, making it easy to implement mutations that apply to multiple program elements, even if they are far apart in the code, such as manipulating the call to a method defined in a different class.
In order to not compromise the consistency of the bytecode model, the application of mutations can depend on the context. 
Finally, transformations are defined in a separate formalism and loaded dynamically.
Therefore, new mutation operators can be easily added without recompiling our tool.

\subsection{A metamodel for Java bytecode}
\label{sec:modbeam}
To facilitate the implementation of mutations as model transformations, we build on the Mod-BEAM approach~\cite{BTNW20,YBRA17}, which stands for ``Modular Bytecode Engineering and Analysis with Models''.
This approach provides an Ecore meta-model capable of fully representing arbitrary Java bytecode.
We briefly summarize the aspects of the meta-model that are most relevant for implementing mutations.

In the meta-model, \ecore{Project} is the root of the meta-model and contains \ecore{Clazz}%
\footnote{This spelling is used avoid confusion with the type \java[basicstyle={\sffamily\scriptsize}]{java.lang.Class}.}
EObjects, each of which represents a \texttt{.class}-file.
These in turn contain nested \ecore{Method} and \ecore{Field} EObjects.
The code of a method is stored as a control flow graph consisting of \ecore{Instruction} EObjects.
 
The abstract EClass \ecore{Instruction} is the root of a type hierarchy that contains a concrete EClass for every possible Java bytecode instruction.
Each instruction type can have specific attributes; for example, an instruction for creating a new object has an attribute \ecore{TypeReference}.
All instructions have in common that they contain a reference to outgoing control flow edges represented by a subtype of the EClass \ecore{ControlFlowEdge}.
This EClass has two properties, namely a reference to the start instruction and a reference to the end instruction.
Possible subtypes are \ecore{UnconditionalEdge}, \ecore{ConditionalEdge} and \ecore{ExceptionalEdge}.

In \cite{BTNW20}, the semantic rules of the bytecode verifier have been specified as OCL~\cite{OMG14} constraints on top of the bytecode meta-model.
Mod-BEAM is implemented as an Eclipse plug-in that can automatically generate instance models for all the compiled code in an Eclipse Java project.
Likewise, the plug-in can automatically generate equivalent \texttt{.class} files from an instance model after it has been transformed.

\subsection{Mutation testing with model transformation}

A mutation of Java bytecode is typically a small change to the bytecode. 
The assumption here is that mutations are applied to a correctly functioning program and that any mutation usually makes the program incorrect.
However, the mutation should not result in a syntax or type error.
Such an error would be detected by the Java Virtual Machine when the class is loaded, thus preventing the execution of the test suite altogether.
Conversely, the changes introduced should be subtle and only detected when a test case uses the correct test data and/or assertion.
For these reasons, mutations usually represent code changes that software developers are likely to encounter.
For example,  it is easy to confuse ``<'' with ``<='' in a termination condition.
A typical example where class-level mutation can be useful is the following: 

\paragraph*{Example}
Developers can easily forget to override a method of a superclass, or misspell a method name so that it accidentally fails to override a method, which can be simulated by a mutation operator by removing the method in the subclass.
To satisfy our goals that mutated programs are still syntactically correct and contain no type errors, the mutation operator must express a condition involving information from two different classes:
A method definition may only be removed from the \emph{child class} if a method of the same signature is defined in the \emph{parent class}.

\begin{figure}
\centering
\includegraphics[width=.8\columnwidth]{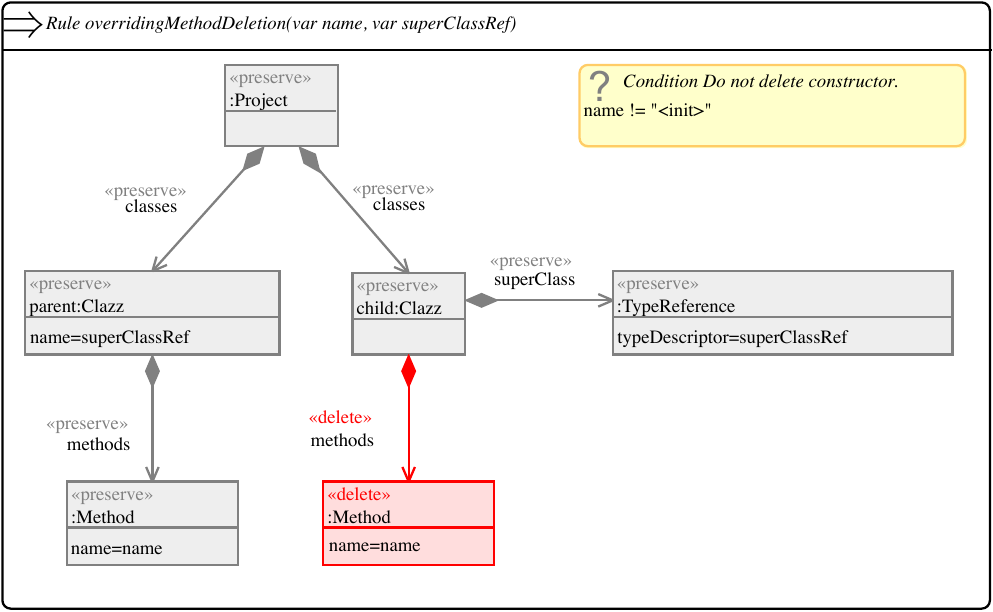}%
\caption{Mutation operator for deleting an overriding method specified as Henshin transformation rule.}%
\label{fig:deleteOverridingMethod}
\end{figure}

Fig.~\ref{fig:deleteOverridingMethod} shows a Henshin rule specifying the deletion of an overriding method in Java bytecode.
It specifies a mutation operator which is applicable in terms of a pattern, represented by the gray model objects with the <<preserve>> stereotype, that must be present in the model:
There are two relevant classes contained in a common project; the class referenced by \texttt{child} has a superclass whose name is referenced by the variable \texttt{superClassRef}.
The name of the second class is the same as the value of the variable \texttt{superClassRef}.
In addition,  both classes contain a method with the name \texttt{name}.%
\footnote{To support overloading, we could also require matching parameter lists, which we omit here for brevity.}
The red object node with the stereotype <<delete>> means that a method with the name \texttt{name} in the subclass is to be removed from the model.
All adjacent references are also deleted. 
The application condition (in yellow) checks that the method name is not \lstinline[language=jvmis]!<init>!, which is reserved for constructors in bytecode.
The mutation operator also has a parameter list that contains the name of the overridden method and the name of the superclass.
In case of an arbitrary mutation, these parameters are automatically set by the rule match.

\begin{figure}
\centering
\includegraphics[width=0.99\columnwidth]{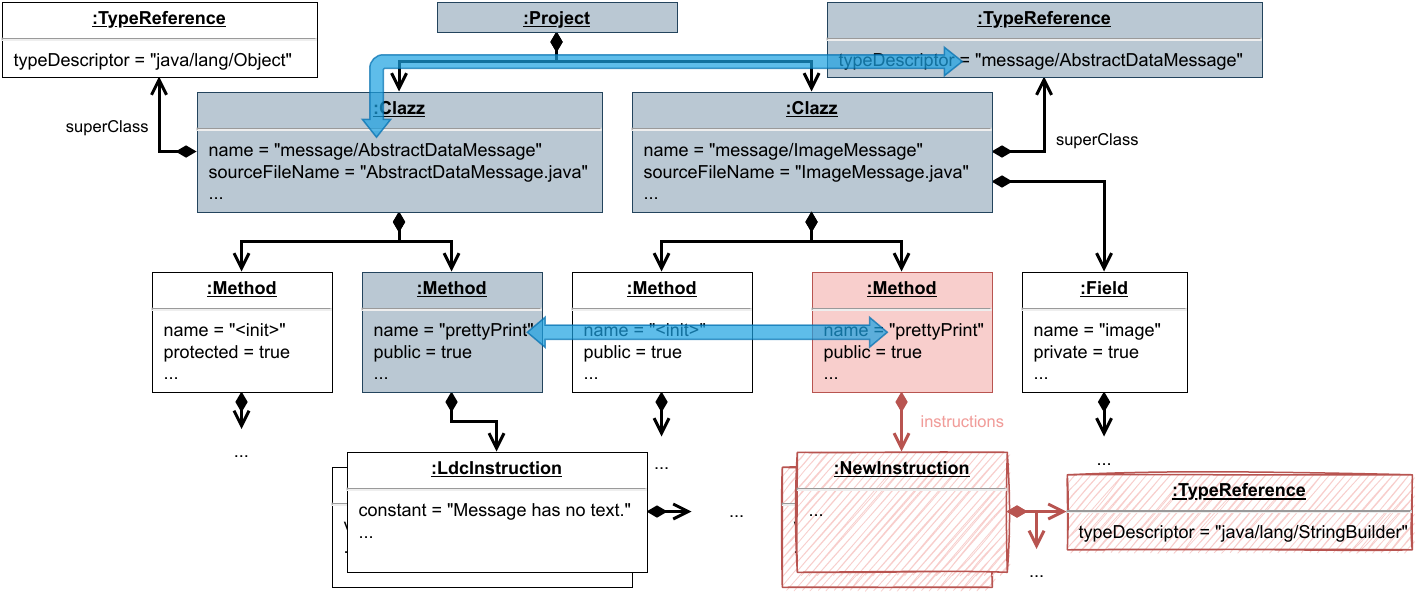}%
\caption{A snippet from the instance model in abstract graphical syntax corresponding to the source code in Fig.~\ref{fig:init-example}, showing the match of the mutation operator in Fig.~\ref{fig:deleteOverridingMethod}.}%
\label{fig:instanceModel}
 \vspace{-.5cm}
\end{figure}

The effect of a concrete application of this mutation operator is illustrated in Fig. \ref{fig:instanceModel}, which shows a snippet from the Mod-BEAM model of the class in Fig.~\ref{fig:init-example}, in an abstract graphical syntax.
The model contains only objects that are directly involved in the mutation that applies the operator in  Fig.~\ref{fig:deleteOverridingMethod}, plus a few more, directly connected objects to show a bit more context.
The objects represented as gray boxes are those that are expected to be present (corresponding to the <<preserve>> stereotype in the Henshin rule). 
The thick blue arrows indicate which attribute values of related objects must match for the mutation operator to apply.
The \texttt{Method} object shown in red is deleted.
The instructions which it contains are red and shaded because they are not explicitly mentioned in the transformation rule but they are detached, turning them into garbage, when their container is deleted.
\medskip

\subsection{Implemented mutation operators}
MMT provides a wide range of mutation operators.
A classical set of mutation operators are arithmetic and relational operator replacements such as replacing ``+'' by ``-'' and ``=='' by ``!=''.  These operators can be typically found in mutation testing approaches.
In addition, we specified class-level mutation operators as presented in\cite{OMK06}.  
This set includes operators that deal with inheritance and polymorphism,  Java-specific properties, and calls of methods in the Java Collection Framework. 
We have also implemented mutation operators for altering method calls of the Java Collection Framework in particular and of any API in general. 
Note that most of these operators have an inverse, which we also implemented.

The more advanced mutation operators are usually defined from the source code perspective, as mutations should stand for errors that developers are likely to make, such as the declaration of a local variable accidentally shadowing a field declaration.
What is a simple line of code inserted by a source code mutation (the variable declaration in the example) may require the mutation of multiple locations in the bytecode (such as replacing all affected field accesses with accesses to the local variable).
This is because the source code is at a more abstract level than the bytecode, and many abstractions are resolved during compilation.
This may also be a reason why several of the advanced mutation operators discussed above are not found in modern bytecode-based mutation testing tools.
We were nevertheless able to implement them in MMT, since Mod-BEAM provides a model at a higher abstraction level than the bytecode itself, leading to the following list:

\begin{enumerate}
\item Data types: change of arithmetic and relational operators 
\item Java-specific: insert/delete \texttt{this}, \texttt{static} or member variable initialization; exchange access methods,  modifier methods,  reference or content comparison 
\item Object orientation: Show/hide subclass variable; delete/insert super or overriding method;change position of overriding method, initialize object with parent/child type, declare member variable/parameter with parent/child type, change cast type to parent/child, delete type cast, exchange body of overloaded method; change arguments in overloaded methods 
\item Library-specific: for Java collections: delete calls of the methods add, clear or remove 
\item for any API: replace parameter/method call; swap parameters/method calls 
\end{enumerate}

When mutating Java bytecode, we want to ensure that the mutated code is still syntactically correct and can be executed.  
Mutation rules can be statically checked for preserving the EMF-constraints~\cite{BET12}. 
If they do so, the resulting mutant is a valid EMF model which can be processed, especially viewed in model editors. 
We checked all 68 implemented mutation operators for this property and found that they all preserve the EMF-constraints. 
Since the bytecode verifier rules were formalized as OCL constraints in \cite{BTNW20}, each model transformation can be accompanied by a post-processing step that checks the corresponding OCL constraints. 
Instead,  a model transformation rule can also be enhanced with application conditions. 
They control the applicability of the transformation rule so that all possible applications lead to transformation results that respect the given set of constraints. 
The construction of such application conditions from a given set of OCL constraints is implemented in an Eclipse-based tool called OCL2AC~\cite{NassarKAT18}.

%% file: 4_tool_support.tex
\section{Tool Support}
\label{sec:toolsupport}

The Model Mutation Testing (MMT) tool, which is an Eclipse plug-in,
can be found at: \url{https://gitlab.uni-marburg.de/fb12/plt/modbeam-mt}.
It supports the management and selection of mutation  operators for concrete testing sessions and also the specification of custom mutation operators for Java bytecode.
For the actual mutation analysis, MMT requires a Java project,  which should also contain JUnit tests, a runtime configuration for executing the test suite, and a set of selected mutation operators. 
The result is displayed in the interactive Mutation Testing View, which contains all mutants, along with information about which mutation operator was applied where in the bytecode, and which mutants were killed.
To specify further mutation operators, Henshin can be used. 
The checks to maintain syntactic correctness are integrated into MMT next.

%% file: 6_related_work.tex
\section{Related work}
\label{sec:relatedwork}

We look at related work in two directions: Other mutation testing tools for Java, in particular for Java bytecode, and other model-driven approaches to mutation testing. 

\subsection{Mutation testing tools for Java}
Javalanche~\cite{schuler2009javalanche}, Jumble~\cite{IPT+07}, LittleDarwin~\cite{littledarwin},  Major~\cite{Jus14},  PITest~\cite{CLH+16},  and Mujava~\cite{OMK06} are existing approaches and tools for mutation testing of Java programs.
Since they were mainly developed for fast mutation testing,  Javalanche, Major and PITest work directly on Java bytecode, only Mujava was developed for Java source code.
All the tools for bytecode mutation testing work only with basic mutation operators, for which it  is easy to guarantee executable programs as output.
Only Mujava and our approach MMT allows to specify more complex mutation operators.
MMT is the only tool that supports complex operators for bytecode, which is possible since mutations are transformations of models and not of trees.

\subsection{Model-driven approaches to mutation testing}
There are two related MDE-based approaches to mutation testing presented in \cite{MBT06} and \cite{GGLM21}.
While \cite{MBT06} presents a mutation testing approach only for model transformation programs, 
a language-independent approach to mutation testing called {\sc Wodel} is presented in \cite{GGLM21}.  
{\sc Wodel} is also a model-based framework for mutation testing implemented as an Eclipse plug-in.
It includes a specification language for mutation operators that has been used, among other specifications,  to specify 77 basic mutation operators for Java source code (not bytecode) such as arithmetic and relational operators. 
{\sc Wodel}  seems well suited for specifying these mutation operators.
To ensure that the specified operators satisfy a certain form of correctness, {\sc Wodel} (as MMT) supports post-processing of mutated models.
Static checks for mutation operators such as the check of EMF constraints are not reported in \cite{GGLM21}.

%% file: 7_conclusion.tex
\section{Conclusion}
\label{sec:conclusion}

This paper presents MMT, a tool for mutation testing of Java bytecode using a model-driven approach to program transformation.
For flexible development of mutation operators, a bytecode model is used that abstracts from actual bytecode.
Mutations of Java bytecode are executed as model transformations.
We implemented MMT as an Eclipse plug-in and specified a set of 68 mutation operators for Java bytecode, which includes not only basic operators such as manipulations of arithmetic operations or relations, but also operators for mutating the class structure, Java-specific properties, and API method calls.
As a next step, we plan to further compare MMT with other mutation testing tools for Java such as Major, Javalanche, LittleDarwin, and {\sc Wodel-Test/Java}. 
Besides studying the weaknesses in test suites that they discover, we will also compare their performance.

%% file: appendix.tex

\onecolumn

\section*{Tool Demo}
\label{sec:toolDemo}

\renewcommand{\floatpagefraction}{0.9}
\renewcommand{\topfraction}{0.95}
\renewcommand\bottomfraction{1.0}
\renewcommand{\textfraction}{0.0}
\renewcommand{\textfraction}{0.0}

The tool demo will present three different workflows of our mutation testing tool MMT.
First, we will demonstrate how to install MMT and explain the basics of mutation testing with MMT.
The second workflow will show the configuration options of the MMT tool.
Finally, we will demonstrate how to extend MMT with custom mutation operators and how to apply them.
In particular, we will present how to specify a mutation operator with Henshin. 

\columnratio{0.35}
\footnotelayout{m}
\begin{paracol}{2}[\subsection*{Workflow 1: Installation and Basic Usage of MMT}]

MMT is an Eclipse plug-in based on the Eclipse IDE for Java and DSL Developer. To install and execute MMT you need Eclipse 2022-09, Java 18 and Maven 3.8 or higher versions.
While some earlier versions may also work, these version numbers correspond to our development environment and are therefore known to  work.
In addition,  the plugins Mod-BEAM\footnote{\url{https://gitlab.uni-marburg.de/fb12/plt/modbeam-mt/modbeam}}  and Henshin\footnote{\url{https://wiki.eclipse.org/Henshin}} are required.
The code to be analyzed by MMT must be provided as an Eclipse Java project.
We have tested MMT on projects with Java up to version 19 and JUnit up to version 5.9.



To show MMT in action, we have chosen a small sample project called \java{WhatsUpApp}, where users can send messages to each other.  
Fig.~\ref{fig:javaSourceCode} shows the first lines of the Java class \java{User}, including the declaration of all fields and the constructor.

\switchcolumn

\vspace{2cm}
\begin{figure}[h]
\centering
\includegraphics[width=0.9\columnwidth]{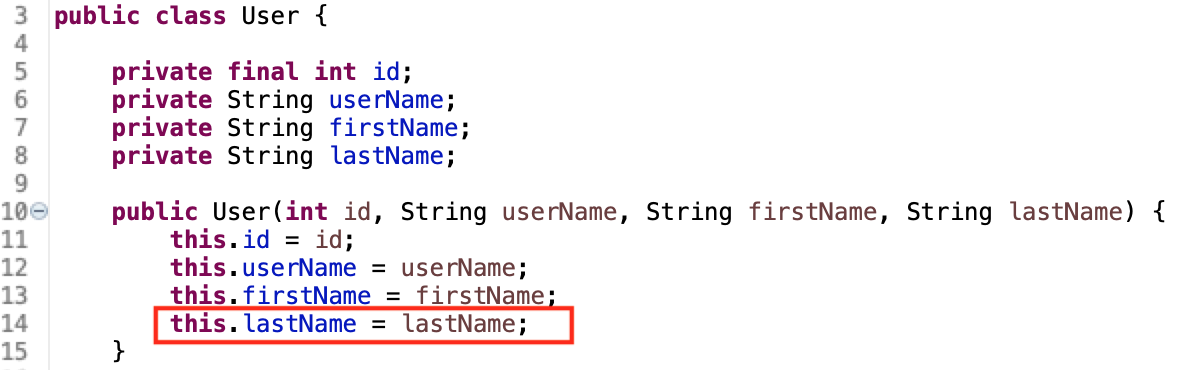}%
\caption{Java source code before mutation. The last field initialization in the constructor of the class ``User'' is to be deleted. }
\label{fig:javaSourceCode}
\end{figure}

\switchcolumn*

It also contains getter and setter methods for all fields (not shown).
The class \java{UserTest} contains several JUnit tests for the class \java{User}; in Fig.~\ref{fig:junitTest} a part of this class is shown, including a test of a getter method and indirectly also of the constructor, since it is used for setup.

\switchcolumn

\begin{figure}[h]
\centering
\includegraphics[width=0.8\columnwidth]{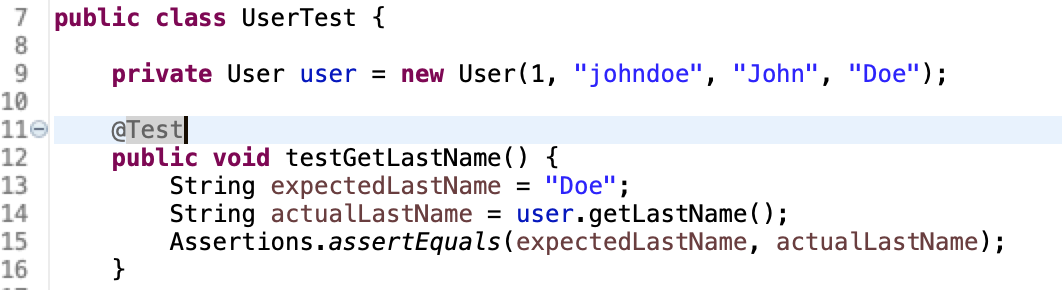}%
\caption{Part of a JUnit test class.}
\label{fig:junitTest}
\end{figure}

\switchcolumn*

When we run these unit tests, we see in Fig.~\ref{fig:junitTestSuccessful} that the class \java{UserTest} has four test cases,  all of which run without errors or failures. 

\switchcolumn

\begin{figure}[h]
\centering
\includegraphics[width=0.5\columnwidth]{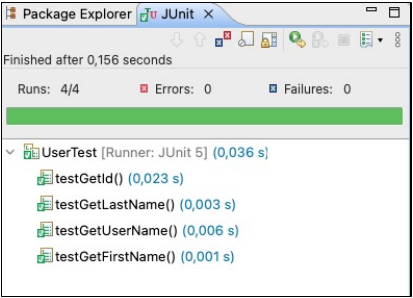}%
\caption{Run of the test class before the mutation. All tests are successful.}
\label{fig:junitTestSuccessful}
\end{figure}

\clearpage
\switchcolumn*

Next, we apply mutation testing to the entire example project.  
The mutation testing view in Fig.~\ref{fig:MutationTestingView} shows which mutation operators have been applied and at which locations, indicated by line numbers in the source code, among other information.
The highlighted entry indicates that the mutation operator fieldVariableInitializationDeletion has been applied to a constructor and that this mutant has been killed.
If a mutant was killed, the rightmost column let's us know which test in particular has killed the mutant.
In addition, the corresponding test case and assertion error are shown in the view (is clipped in this figure for brevity).
Otherwise, MMT view simply informs us that the mutant was not detected by the test suite. 

\switchcolumn

\begin{figure}[h]
\centering
\includegraphics[width=0.9\columnwidth]{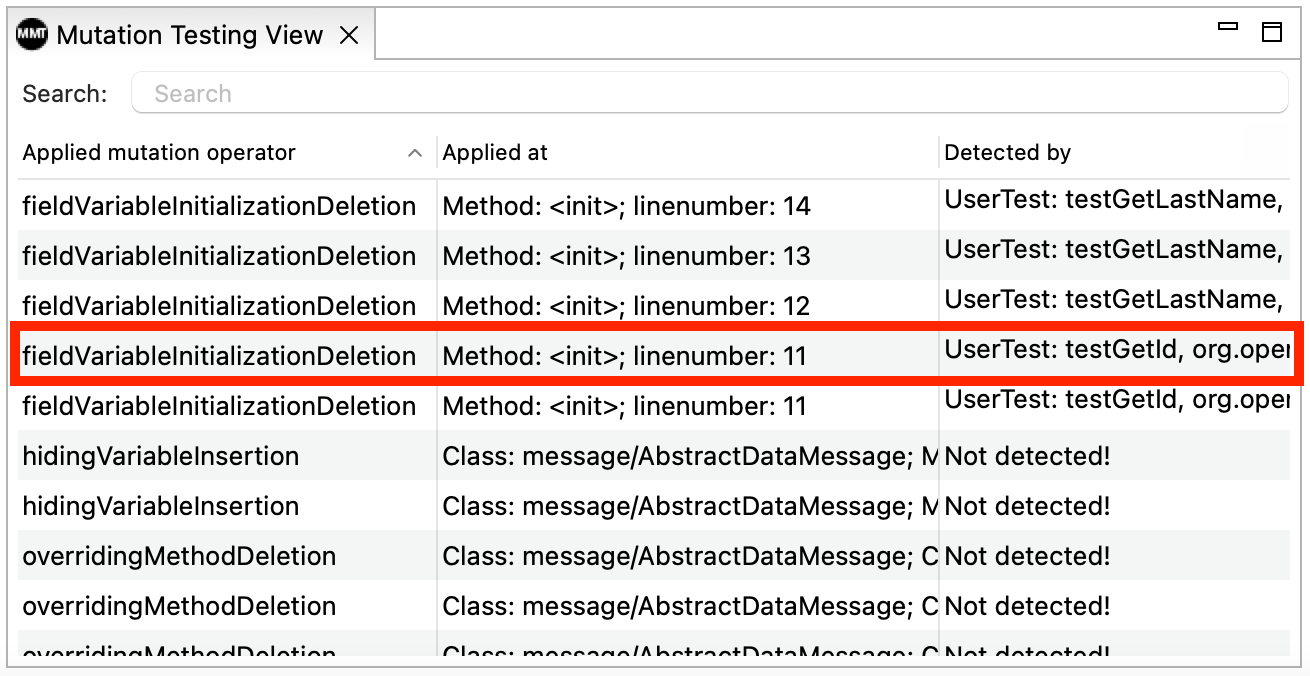}%
\caption{The Mutation testing view.}
\label{fig:MutationTestingView}
\end{figure}

\switchcolumn*

Double-clicking an entry in the list will display details for the selected mutant.
Fig.~\ref{fig:MutationTestingViewDetail} shows the detail dialog for the selected mutant.
We see that the mutation operator \emph{fieldVariableInitializationDeletion} has been applied at line 11 of a constructor method.
The mutant was killed by the test \java{testGetId} in the test class \java{UserTest}.
Fig.~\ref{fig:MutationTestingViewDetail-live} shows the detail dialog for another mutant that was not killed by the test.
We see a warning that the mutant is live along with the information that the mutation operator \emph{fieldVariableInitializationDeletion} has been applied again in a constructor, but this time at line 15.

\switchcolumn

\begin{figure}[h]
\centering
\includegraphics[width=0.7\columnwidth]{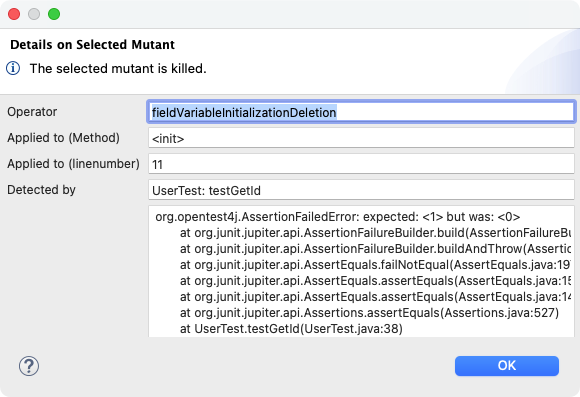}%
\caption{Detail dialog for selected killed mutant.}
\label{fig:MutationTestingViewDetail}
\end{figure}

\begin{figure}[h]
  \centering
    \includegraphics[width=0.7\columnwidth]{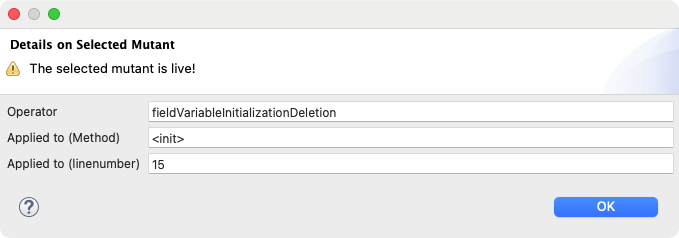}
\caption{Detail dialog for selected live mutant.}
\label{fig:MutationTestingViewDetail-live}

\end{figure}

\clearpage
\switchcolumn*[\subsection*{Workflow 2: Configuration of MMT}]

Pressing the MMT button on the Eclipse toolbar displays the Mutation Testing Wizard (see Fig.~\ref{fig:mutationOperators}), where the user can select the Java project to be tested, configure the run configuration for the tests, and select the mutation operators to be applied. 

\switchcolumn

\begin{figure}[h]
\centering
\includegraphics[width=0.9\columnwidth]{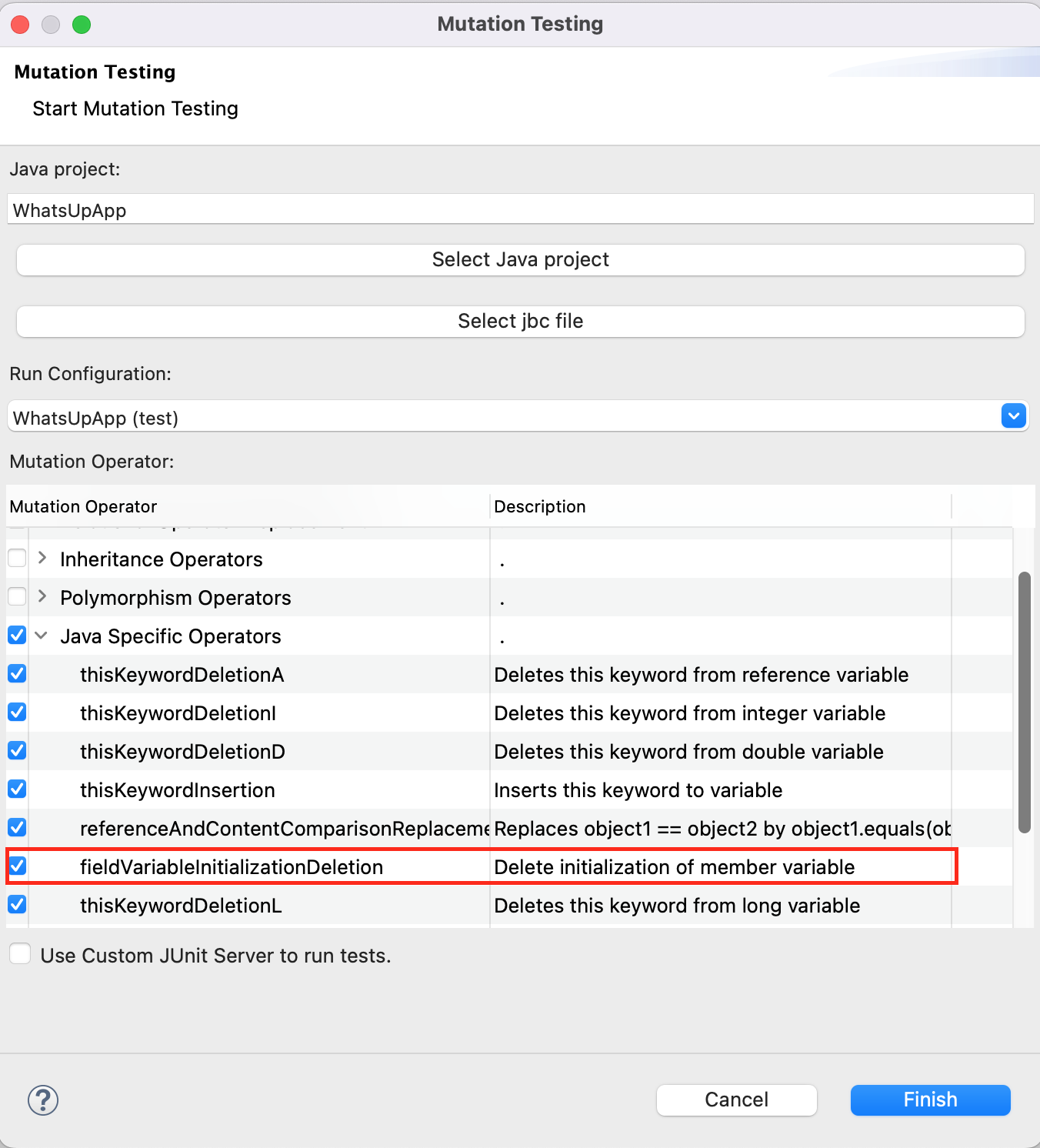}%
\caption{Selection of mutation operators in MMT.}
\label{fig:mutationOperators}
\end{figure}

\switchcolumn*
To use custom mutation operators,  they have to be specified with an external model transformation tool that is based on EMF; currently, only Henshin is supported, but integration with other model transformation tools is straightforward and will be possible in the future.
The custom Henshin files can be added in the MMT preferences view (see Fig.~\ref{fig:MMTPreferences}); after that they are also selectable mutation operators in the mutation testing wizard.

\switchcolumn

\begin{figure}[h]
\centering
\includegraphics[width=0.9\columnwidth]{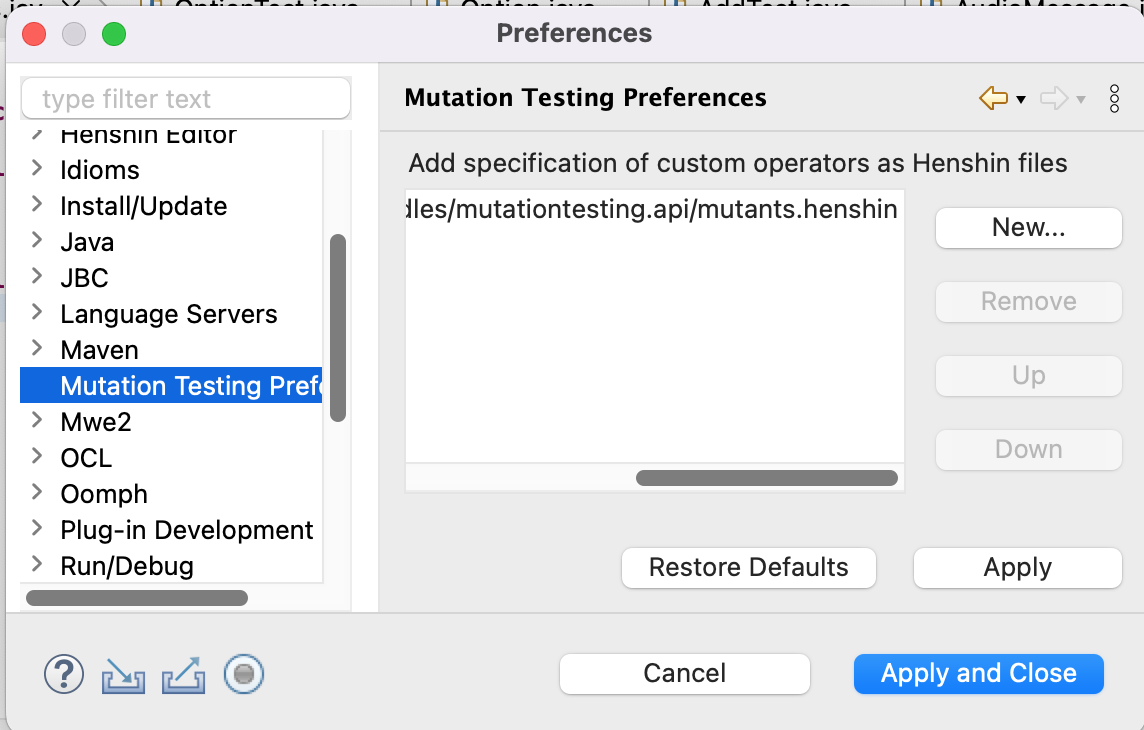}
\caption{Preferences view of MMT.}
\label{fig:MMTPreferences}
\end{figure}

\clearpage
\switchcolumn*[\subsection*{Workflow 3: Extending MMT with a Custom Mutation Operator}]

As an example of specifying a custom mutation operator, we demonstrate how to develop a Henshin transformation rule for the operator \java{fieldVariableInitializationDeletion}.
Note that mutation operators in MMT are specified at the bytecode level. 
To understand what part of the bytecode needs to be changed to delete the initialization of a field variable, we show a part of the bytecode of the class \java{User} before and after the mutation in Fig.~\ref{fig:BytecodeBeforeAfter}.  

\switchcolumn

\begin{figure}[h]
\centering
\includegraphics[width=0.9\columnwidth]{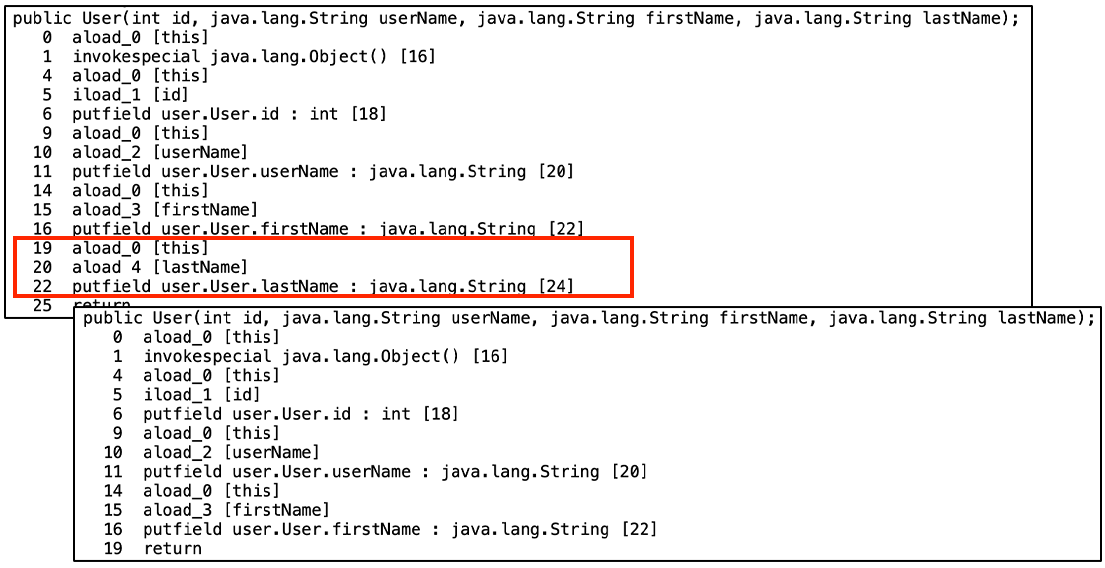}%
\caption{Bytecode of class \java{User} before and after a mutation that deletes the initialization  of the field variable \java{lastname}. The instructions to be deleted are highlighted by a red rectangle. }
\label{fig:BytecodeBeforeAfter}
\end{figure}

\switchcolumn*
The bytecode model corresponding to the left bytecode is shown in Fig.~\ref{fig:BytecodeModel}. 
The EObjects of type \java{Instruction} highlighted here correspond to the instructions highlighted in the previous figure. 

\switchcolumn

\begin{figure}[h]
\centering
\includegraphics[width=0.6\columnwidth]{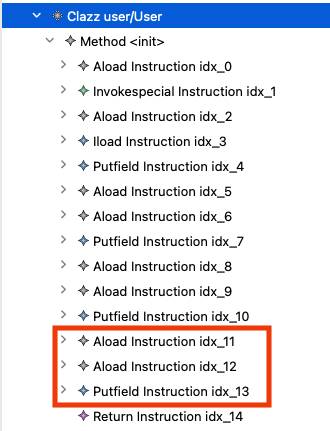}%
\caption{Bytecode model for the upper part of the bytecode on the left in Fig.~\ref{fig:BytecodeBeforeAfter}}
\label{fig:BytecodeModel}
\end{figure}

\clearpage
\switchcolumn*

Fig.~\ref{fig:exMutationOperator} shows a Henshin rule specifying the mutation operator \java{fieldVariableInitializationDeletion}.
We see that three instructions are deleted,  the last one of which is a \java{PutfieldInstruction}. 
The first one must be an \java{AloadInstruction}. 
This ensures that an initialization is only deleted when the field variable is set with a parameter value of the constructor.
The reference to the field, its descriptor and the reference to the declaring class are also deleted. 
This operator works only if the type of the field is a basic data type (as required by the attribute condition).
Applying this operator to the bytecode on the left in Fig.~\ref{fig:BytecodeBeforeAfter}, we get the bytecode on the right in the same figure. 

\switchcolumn

\begin{figure}[h]
\centering
\includegraphics[width=0.9\columnwidth]{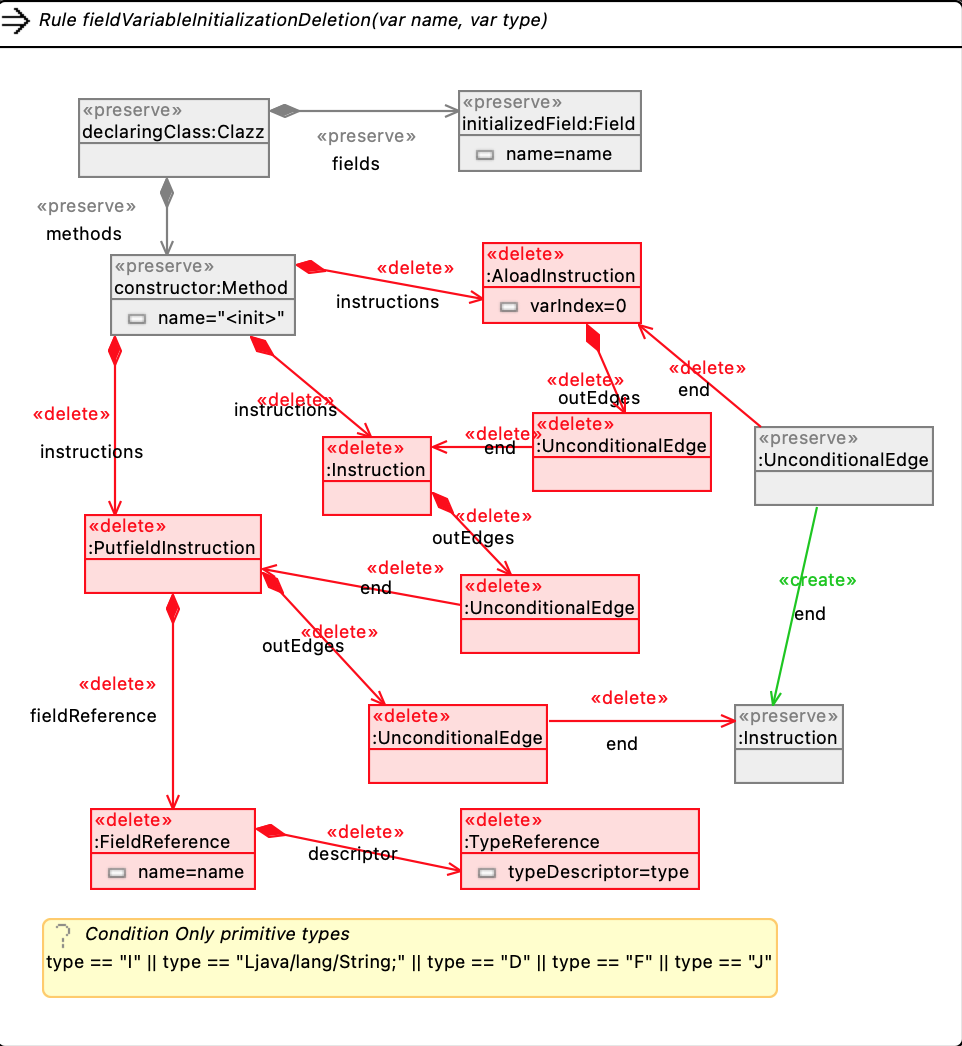}%
\caption{Mutation operator for deleting a field initialization if the field has a primitive type, specified as Henshin rule.}
\label{fig:exMutationOperator}
\end{figure}

\switchcolumn*
A second example of mutation is the deletion of an overriding method. 
This can be a subtle change in the code that can easily lead to unexpected behavior.
To understand what part of the bytecode needs to be changed to delete an overriding method, we show part of the bytecode of the class \java{ImageMessage}, which extends the class \java{AbstractDataMessage}. 
The code before the mutation is shown in Fig.~\ref{fig:OverridingBefore}, the beginning of the class \java{ImageMessage} is shown as source code and bytecode model.  

\switchcolumn 

\begin{figure}[h]
\centering
\includegraphics[width=0.9\columnwidth]{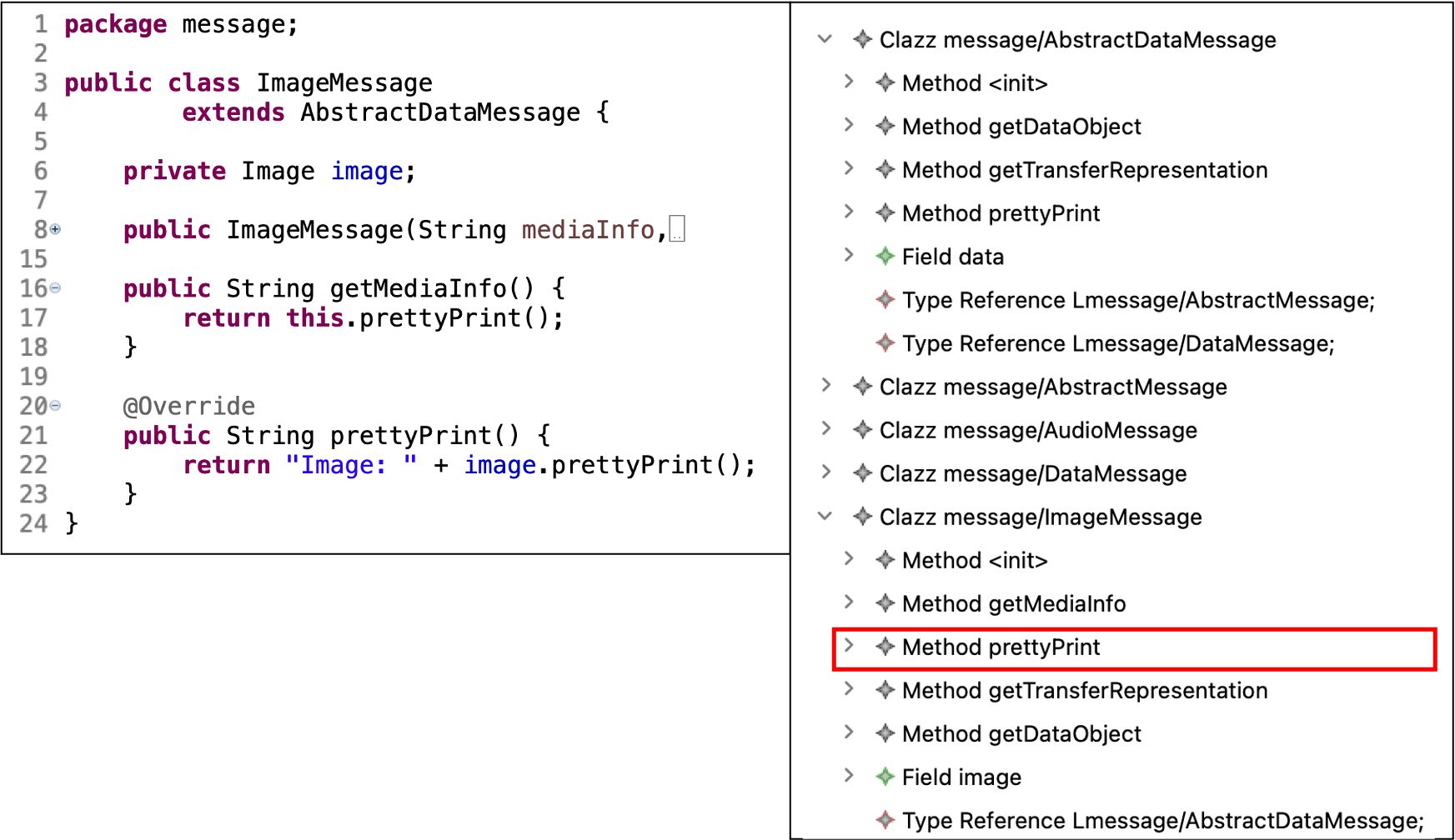}%
\caption{Snippet of source code and corresponding Bytecode model part to illustrate a mutation that deletes an overriding method}
\label{fig:OverridingBefore}
\end{figure}

\clearpage
\switchcolumn*

The overriding method to be deleted is marked by a red rectangle.
Fig.~\ref{fig:OverridingAfter} shows the bytecode model after the mutation where this overriding method was deleted.

\switchcolumn

\begin{figure}[h]
\centering
\includegraphics[width=0.6\columnwidth]{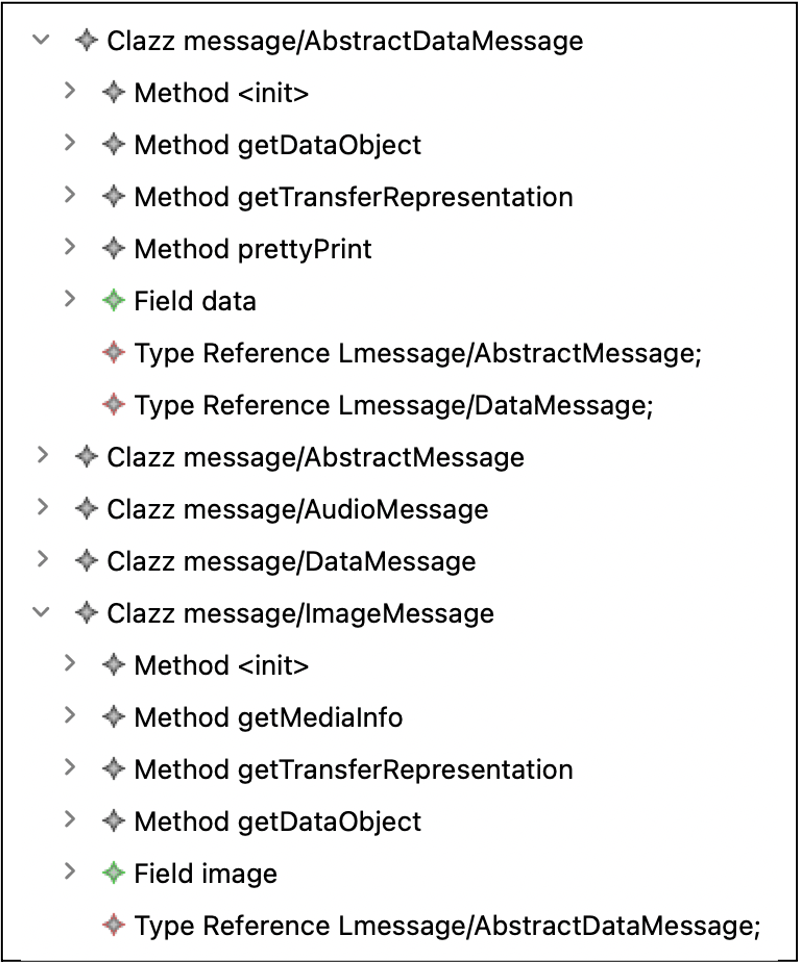}%
\caption{Bytecode model part after a mutation that deletes an overriding method}
\label{fig:OverridingAfter}
\end{figure}

\switchcolumn*

Fig.~\ref{fig:OverridingRule} shows a Henshin rule specifying the mutation operator \java{overridingMethodDeletion}.
We can see that a method is deleted along with all of its instructions.
There is also an application condition that ensures that the method to be deleted is not a constructor.
This is important because constructors are treated as methods with the reserved name \lstinline[language=Bytecode]!<init>!, but overriding works differently for constructors.
Even if the superclass generally also has a method named \lstinline[language=Bytecode]!<init>! in bytecode, this constructor method will not be inherited if the constructor is missing in the subclass.
Applying this operator to the bytecode model part on the right of Fig.~\ref{fig:OverridingBefore}, we essentially get the bytecode model in  Fig.~\ref{fig:OverridingAfter}.

\switchcolumn

\begin{figure}[h]
\centering
\includegraphics[width=0.9\columnwidth]{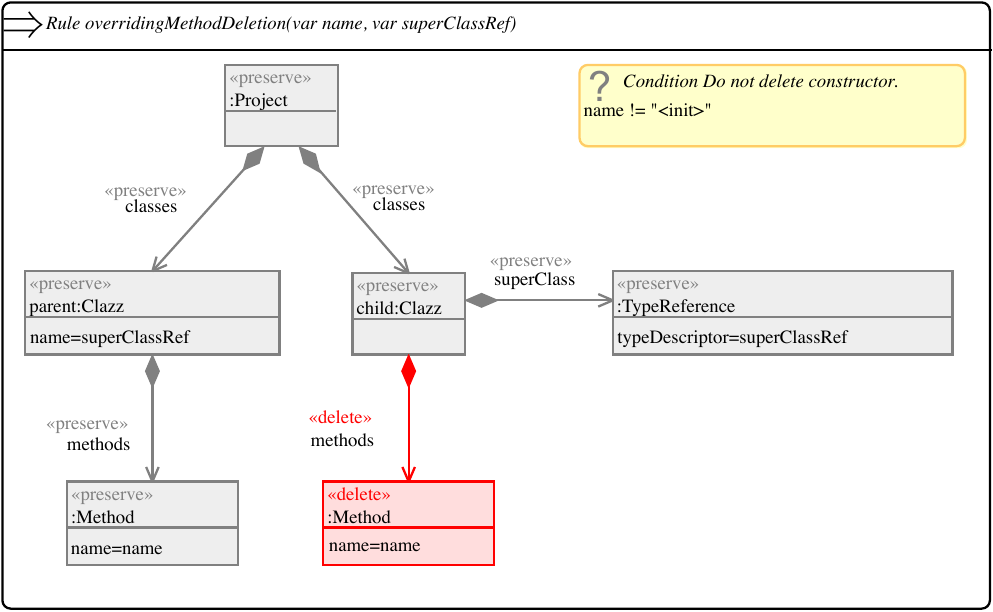}%
\caption{Mutation operator for the deletion of an overriding method}
\label{fig:OverridingRule}
\end{figure}

\clearpage
\switchcolumn*

Adding custom mutation operators typically results in additional mutants that may be difficult for a test suite to detect, and thus may reveal  additional weaknesses in the test suite.
We can see this when we run a mutation testing process on our example project without and with the new mutation operator \emph{overridingMethodDeletion}.
Fig.~\ref{fig:mutation-score-without} shows the dialog where MMT summarizes the result of the mutation test run with all predefined mutation operators.
The mutation score is 0.83, which means that 83\% of the mutants have been detected.
Fig.~\ref{fig:mutation-score-with} shows the same dialog but after mutation testing with the new mutation operator, additionally.
In this second run, the mutation score dropped to 69\%.

\switchcolumn

\begin{figure}[h]
\centering
\includegraphics[width=0.9\columnwidth]{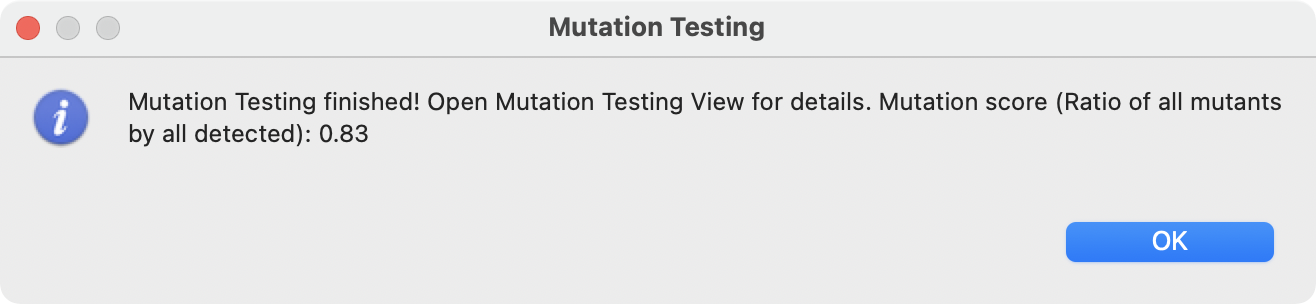}%
\caption{Result summary of mutation testing \emph{without} the mutation operator \emph{overridingMethodDeletion}.}
\label{fig:mutation-score-without}
\end{figure}

\begin{figure}[h]
\centering
\includegraphics[width=0.9\columnwidth]{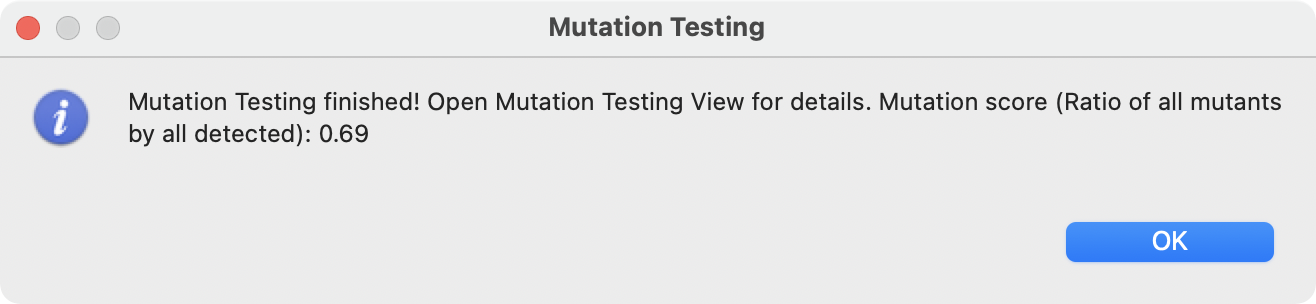}%
\caption{Result summary of mutation testing \emph{with} the mutation operator \emph{overridingMethodDeletion}.}
\label{fig:mutation-score-with}
\end{figure}

\switchcolumn*

Inspecting the new mutants, we see that a mutant resulting from the deletion of the method \java{prettyPrint} in \java{ImageMessage} could not be detected by the test suite (see Fig.~\ref{fig:MutationTestingViewDetail2}).
This shows that we need an additional test that checks if calling this method really returns the specific result expected from an image message. 

\switchcolumn

\begin{figure}[h]
\centering
  \includegraphics[width=0.7\columnwidth]{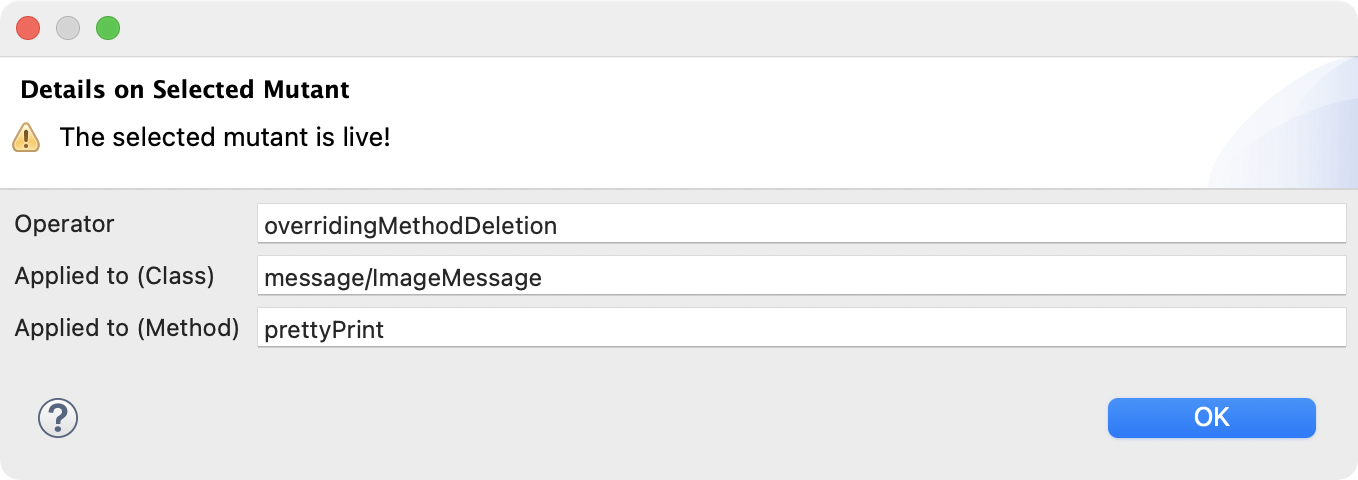}
  \caption{Details showing that the mutation operator \emph{overridingMethodDeletion} produced a live mutant.}
  \label{fig:MutationTestingViewDetail2}
\end{figure}

\clearpage
\switchcolumn*

If we add the test in Fig.~\ref{fig:testOverridingMethod} and run MMT again, we see that the mutation score increases again (see  Fig.~\ref{fig:mutation-score-with-additional-test}) to 79\%  and that the same mutant is now killed (see Fig.~\ref{fig:mutant-killed-with-new-test}).
The test suite still passes when run against the original version of the code, but since the assertions are now stronger, we are more likely to detect regressions in the future if bugs are introduced into the code.
\textbf{\textit{With the MMT approach, it was therefore possible to improve the quality of the test suite.}}

The exercise of adding a stronger test case can even help us identify bugs directly, because we are forced to understand the contract of the method under test better.
In the example presented here, we can look at the requirements of the pretty-printing message for an image message to see what its textual representation should look like.
Let's assume that the requirements state that the textual representation should be the word \java{"Image"} followed by a URL from which the image was loaded.
If we write a test that checks the method result for exactly this format, we would discover that the method \java{prettyPrint} in the class \java{ImageMessage} wrongly includes the inappropriate dummy message returned by the super implementation, instead of the source URL.

\switchcolumn

\begin{figure}[h]
\centering
  \includegraphics[width=0.9\columnwidth]{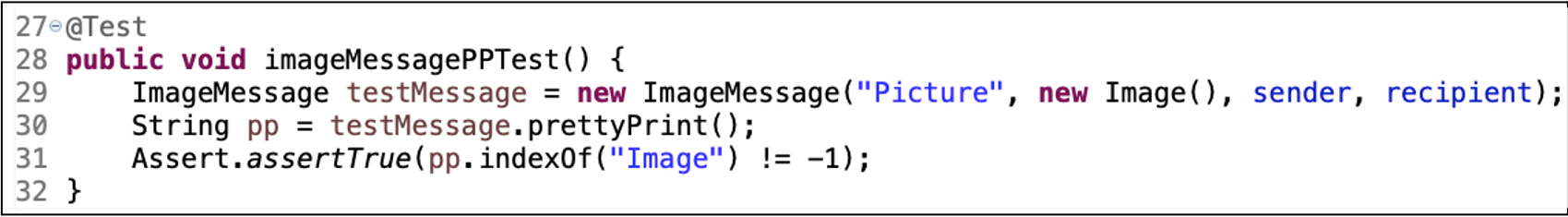}
  \caption{Additional test for checking specific result of \java{prettyPrint} in \java{ImageMessage}.}
  \label{fig:testOverridingMethod}
\end{figure}

\begin{figure}[h]
\centering
  \includegraphics[width=0.7\columnwidth]{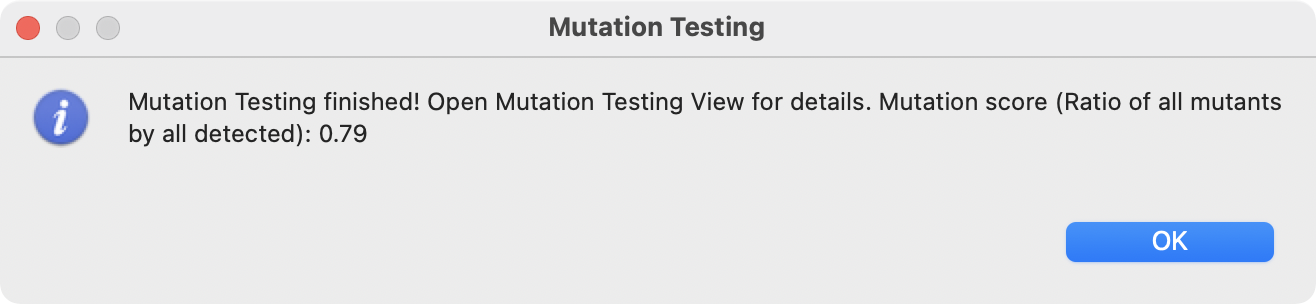}
  \caption{Details showing that the mutation operator \emph{overridingMethodDeletion} produced a live mutant.}
  \label{fig:mutation-score-with-additional-test}
\end{figure}

\begin{figure}[h]
\centering
  \includegraphics[width=0.7\columnwidth]{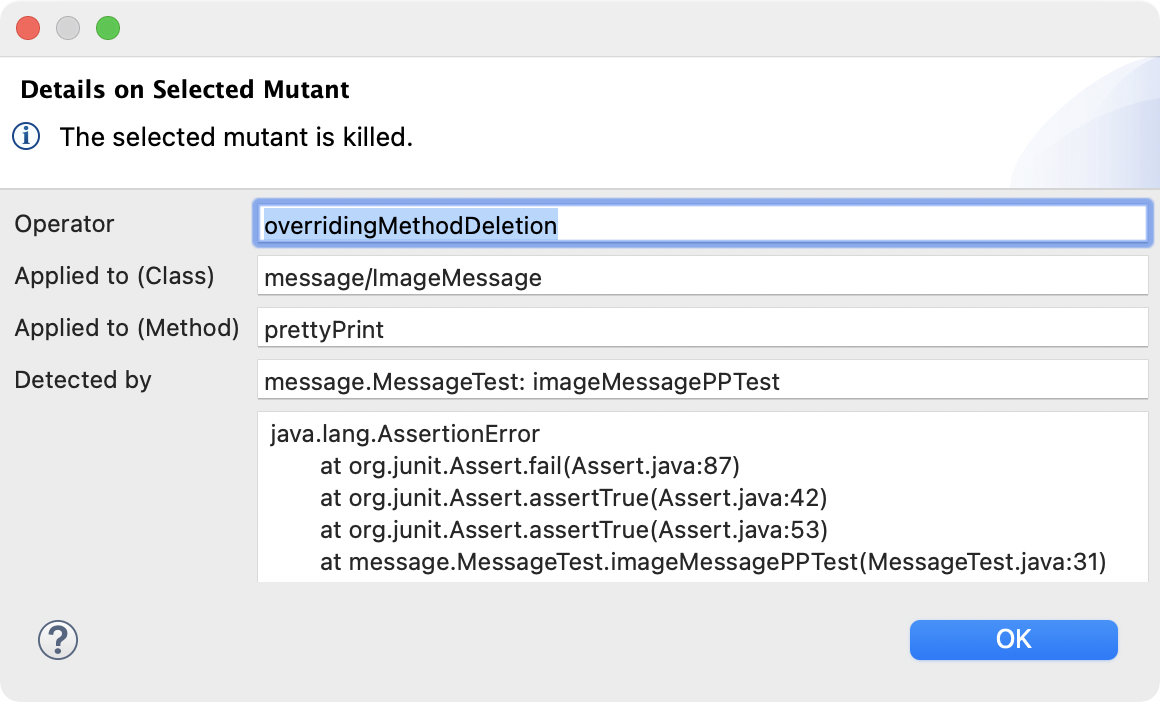}
  \caption{Result summary of mutation testing \emph{with} the mutation operator \emph{overridingMethodDeletion} and with additional test.}
  \label{fig:mutant-killed-with-new-test}
\end{figure}

\end{paracol}